\documentclass[a4paper,11pt,reprint,onecolumn,longbibliography,eqsecnum,floats,aps,amsmath,amssymb,nofootinbib,prd,showpacs,notitlepage]{revtex4-1}
 \usepackage[a4paper,top=3cm,bottom=3cm,left=2cm,right=2cm,marginparwidth=1.75cm]{geometry}
\usepackage[toc,page]{appendix}
\usepackage[english]{babel}
\usepackage[utf8]{inputenc}
\usepackage{amsmath}
\usepackage{amssymb}
\usepackage{lmodern}
\usepackage{color}
\usepackage{nccmath}
\usepackage{graphicx}
\usepackage{wrapfig}
\usepackage{enumerate}
\usepackage{hyperref}
\usepackage{braket}
\usepackage{float}
\usepackage{dsfont}
\usepackage{mathrsfs}  
\usepackage{natbib}

\newcommand{\id}{\mathbb{I}}

\usepackage{indentfirst}

\DeclareMathAlphabet{\dutchcal}{U}{dutchcal}{m}{n}
\SetMathAlphabet{\dutchcal}{bold}{U}{dutchcal}{b}{n}
\DeclareMathAlphabet{\dutchbcal} {U}{dutchcal}{b}{n}
\begin{document}

\title{Rainbow Oppenheimer-Snyder collapse and the entanglement entropy production.}

\author{Micha{\l} Bobula}
	\email{michal.bobula@uwr.edu.pl}
	\affiliation{Institute for Theoretical Physics, Faculty of Physics and Astronomy, University of Wroc{\l}aw, pl. M. Borna 9, 50-204  Wroc{\l}aw, Poland}
\author{Tomasz Paw{\l}owski}
	\email{tomasz.pawlowski@uwr.edu.pl}
	\affiliation{Institute for Theoretical Physics, Faculty of Physics and Astronomy, University of Wroc{\l}aw, pl. M. Borna 9, 50-204  Wroc{\l}aw, Poland}

\begin{abstract}
The dust ball collapse is studied in the context of the "rainbow metric" approach (where the matter content is supplemented with a scalar field perturbation) to the Oppenheimer-Snyder collapse scenario within the framework of loop quantum cosmology. The global spacetime structure is determined for this scenario and subsequently used to evaluate the entanglement entropy via a slight adaptation of existing formulas. The resulting model is shown to qualitatively resemble the Reissner-Nordstr\"om black hole spacetime, in particular, it still contains singularities, which allows us to define the entropy only for portions of the null infinity. The consequence of these results for the black hole information loss paradox is discussed. Furthermore, the results in presence of the scalar field are used in discussion regarding the viability of the selected scenario, in particular, the assumption of stationarity used to determine the exterior metric.
\end{abstract}

\pacs{04.60.Kz, 04.60.Pp, 98.80.Qc}

\maketitle

\section{Introduction}

\subsection{Motivation}

The existence of singularities in general relativity is a widely accepted yet troublesome prediction, especially when one tries to include in the physical picture the quantum aspects of reality. Infinite curvatures usually featured at such make them impassable barriers to making deterministic predictions at the classical level already. The situation becomes even worse once one tries to reconcile GR with other fundamental laws of physics. That is particularly visible in the process of black hole formation. Notably, one of the most challenging puzzles related with this process is the black hole information loss paradox -- the description using a formalism of quantum fields on fixed black hole background indicates a breakdown of predictability of the theory \cite{Hawking:1976ra} -- the violation of the unitarity of the quantum fields evolution. However, the history of the development of physical models/theories teaches us, that such deficiencies are usually indicators of the limits to the descriptions applied rather than fundamental features of physical reality. For that reason, it is generally believed that a more advanced prescription, possibly reconciling the principles of general relativity with quantum physics will be free from said deficiencies. Such description would then most likely change the causal structure of a spacetime describing a black hole formation and evaporation process.

One of the approaches in building these descriptions is the one, where the spacetime geometry and the non-gravitational fundamental interactions are treated on the same,  namely \textit{quantum} footing. One attempt at building such a theory is Loop Quantum Gravity (LQG), built around a central principle of strict background independence of fundamental components of physical systems description. While its mathematical foundations are well established, the yet unresolved ambiguities and the very high level of technical complication have made it next to impossible to extract exact quantum dynamical predictions in essential physical scenarios. For that reason, the community exploited a multitude of its simplifications. One (probably most widely studied) is symmetry-reduced variants of the full framework tailored for specific classes of geometries -- a prominent example here is Loop Quantum Cosmology (LQC). The achievement of such is a replacement of big bang singularity with \textit{quantum bounce} \cite{Ashtekar:2006rx}. That may suggest the analogous phenomenon in the black hole interior. There, quantum gravity effects should prevent the formation of black hole singularities. Instead, the collapsing matter is expected to bounce and possibly reemerge in a white-hole-like region. This constitutes the so-called \emph{black-to-white hole transition} paradigm \cite{Stephens:1993an, Ashtekar:2005cj}. The removal of singularities drastically changes the spacetime global structure -- in principle, an observer located at future infinity can reconstruct the entire spacetime, thus resolving the information loss paradox by restoring the unitarity of the evolution. However, figuring out whether (and how) this paradigm is realized requires probing an actual gravitational collapse scenario within the considered framework (be it an extension to LQC or a more involved one).

\subsection{Gravitational collapse in polymer quantization}

So far, several approaches to formulating a consistent quantum description of black holes within LQG-related frameworks have been pursued and discussed in the literature. This includes the so-called reduced phase space quantization (in the form of several distinct approaches) that has been widely applied to black-hole-related scenarios. These are based on the fact that in GR there exists a wide class of black hole spacetime descriptions admitting spatial symmetries, further implementing an idea of performing the symmetry reduction of a given model the quantization. 
Probably, the most known and studied (classical) candidate for this scheme is Schwarzschild space-time, where the spherical symmetry allows to reduce the description to a $1+1$ dimensional one. The particular construction of the framework has been performed in several ways. 

One of them is the so-called midi-superspace approach \cite{Bojowald:2004af, Bojowald:2005cb}, where the dimensional reduction is based specifically on spherical symmetry and the abovementioned reduction to $1+1$ system, which is quantized then per analogy with full LQG with quantum geometry states represented by the spin networks now living on 1-dimensional graphs (chains). There, the resulting (reduced) constraint algebra is a true Lie algebra  \cite{Gambini:2008dy, Gambini:2013ooa, Gambini:2020qhx}. The existing works implementing this scenario suggest a modified global spacetime picture resembling the classical Reissner-Nordstr\"om one, but without its singularities \cite{Gambini:2008dy, Gambini:2013ooa}. 

Another approach taken is the use of isometry between homogeneous cosmological Kantowski-Sachs spacetime and the Schwarzschild black hole interior, allowing (due to the homogeneity of the former) the quantization via a direct application LQC tools \cite{Modesto:2005zm, Boehmer:2007ket, Corichi:2015xia, Olmedo:2017lvt}. However, these problems generically suffer from the fact that the embedding of Kantowski-Sachs spacetime in the Schwarzschild one is singular, which led to spurious quantum gravity effects at the black hole event horizon, making the process of extending the resulting geometry to the black hole exterior difficult. To resolve this problem, several phenomenological modifications to the LQC formalism have been proposed, however, due to their ad-hoc nature and violation of certain properties considered essential in LQC (well-definiteness of the theory in the so-called fiducial cell expansion limit) they are far from satisfactory. 

A much more successful variation of this approach is based on the property, that in symmetry-reduced spherically symmetric spacetime, being a $1+1$ dimensional system, the roles of radius and time could be interchanged. This allowed to construct a quantum desription of a static black hole exterior that is an analog of the LQC one \cite{Ashtekar:2018cay}, thus giving a consistent quantum picture. Even there however one observes the discrepancies for classical GR limit in the form of a modification by an effective force decaying with radius slower than the Newtonian one. This modification, while extremely weak, will become dominant as one approaches spatial infinity, thus possibly affecting the existing (and already experimentally tested) framework of gravitational waves.

Beyond static/stationary spacetimes the reduced phase space quantization can also be applied to Lemaître– Tolman–Bondi class of spacetimes \cite{Kelly:2020lec, Kelly:2020uwj, Han:2020uhb, Husain:2022gwp}. In particular, one can modify the classical LTB spacetime dynamics by imposing phenomenological LQC-inspired effective dynamics modifications (holonomy corrections). Notably, such modifications induce bouncing Oppenheimer-Snyder collapse scenario \cite{Kelly:2020lec, Kelly:2020uwj}. Alternatively, one can quantize an inhomogeneous classical model by introducing the discretization of the spacetime coordinates. Then, each node of the constructed lattice is subject to the LQC quantization procedure, and the resulting total Hilbert space is the tensor product of Hilbert space at each node. The outcome of the employed framework is the singularity-free, effective spacetime \cite{Husain:2022gwp}.

Yet another category of approaches is the spin foam (covariant LQG) approach to the black-to-white hole transition \cite{Haggard:2014rza,DAmbrosio:2018wgv, Bianchi:2018mml,DAmbrosio:2020mut, Soltani:2021zmv}. Unlike in the canonical approaches discussed above, there one implements the (simplifications of the) analog of path integral description -- the spin foam formalism \cite{Perez:2012wv, Rovelli:2014ssa}. The evolution of the system is captured via transition amplitudes. The general strategy is to start with an effective metric continuing from
the trapped to the anti-trapped region of the black-hole-like/white-hole-like interior (in a certain limit, the metric becomes the interior of the Schwarzschild black hole). Next, Kruskal-like geometry is glued to the former. However, that procedure is non-trivial for subregions of expected high Planck regime. Consequently, spin foam amplitudes are utilized to determine the dynamics there. The resulting global picture describes a black-to-white hole transition.

In this work, we focus on the so-called \textit{rainbow metric} approach \cite{Ashtekar:2009mb, Assanioussi:2014xmz, Parvizi:2021ekr} to Oppenheimer-Snyder collapse scenario. Its main idea is to consider the evolution of quantum fields in (symmetric) quantum space-time, from which the semi-classical effective background metric emerges (is derived). In the context of our work, this matter field is represented by a massless scalar field. The resulting effective geometry is mode-dependent, that is each mode of the scalar field induces a different effective (semi)classical metric (hence the name \textit{rainbow}). In particular, one can examine quantum dynamics of gravitational degrees of freedom (quantized according to LQC) coupled to the scalar field quantized according to the Schrödinger picture. The Schrödinger-like equation describes the evolution of the total system, which can be solved with semiclassical methods. Then, the resulting approximate evolution of the system (accounting for the back-reaction of background geometry) is identified with the evolution of standard QFT living on \textit{some} classical background space-time (also called as \textit{rainbow metric} or \textit{dressed background metric}). In this way, the effective mode-dependent geometry emerges in the formalism. 

\section{Rainbow metric approach to Oppenheimer-Snyder collapse}

In this section, we introduce a model of the loop quantum corrected Oppenheimer-Snyder collapse scenario. We consider both pure dust ball collapse and a similar scenario accounting for massless scalar field perturbations. The interior of the dust ball, being flat and homogeneous is quantized via LQC methods. We obtain effective, bouncing Friedmann–Lemaître–Robertson line element for that region. However, we do not quantize the exterior. Instead, we assume a general form of a black-hole-like metric there -- an Eddington-Finkelstein line element with the unspecified metric component -- and the exact form of such will be determined from the junction conditions on the boundary surface (in this case, the surface of the collapsing ball). Finally, we are able to construct a conformal diagram  (the so-called \emph{Penrose diagram}) for the resulting space-time.

\subsection{Dust ball interior}

In order to model the interior of the collapsing dust ball, we follow the formalism developed in \cite{Ashtekar:2009mb, Husain:2011tm, Husain:2011tk, Assanioussi:2014xmz, Parvizi:2021ekr}. A classical action describing the interior consists of gravitational degrees of freedom coupled to matter fields: the dust field $T$ and the massless scalar field $\phi$. A flat FRW line element is imposed for the background geometry, and then the Arnowitt-Deser-Misner (ADM) formulation is constructed. The canonical quantization is performed in a hybrid way being a standard in LQC: polymer quantization is adapted for the gravity sector, while the matter fields are quantized in standard Schrödinger representation. The theory is further deparametrized with respect to the dust field $T$ (playing the role of \textit{internal time}). In that case, the kinematical Hilbert space becomes a physical one directly \cite{Husain:2011tk}. Subsequently, we can write a Schrödinger-like equation (originating from the Hamiltonian constraint) describing the evolution of a state living in kinematical Hilbert space $\Psi \in \mathcal{H}_{\mathrm{grav}} \otimes \mathcal{H}_T \otimes \mathcal{H}_\phi $ (where $\mathcal{H}_\text{grav} $ is the gravitational Hilbert space) as
\begin{equation}
\label{evolution}
  i \hbar \partial_T \Psi(\dutchcal{v}, \phi, T)
  = \left(\hat{H}_{\text {grav }}\otimes\id +\id\otimes \hat{H}_\phi\right) \Psi(\dutchcal{v}, \phi, T) \, ,
\end{equation}
where the wave function $\Psi(v,\phi,T)$ is the decomposition of the physical state ``at the moment'' $T$ with respect to the (generalized) eigenstates $|\dutchcal{v}\rangle$ of the oriented volume (for the geometry subspace) and field $(\phi|$ (for the matter) operators respectively. 
The gravitational Hamiltonian is of the form of a relatively simple composite operator
\begin{equation}
  \hat{H}_{\text {grav }}=-\frac{3 \pi G \hbar^2}{2 \alpha_o} \sqrt{|\hat{\dutchcal{v}}|} \sin ^2(\hat{\dutchcal{b}}) \sqrt{|\hat{\dutchcal{v}}|} \, ,
\end{equation}
being a function of:
\begin{enumerate}[(i)]
  \item the rescaled oriented volume operator $\hat{\dutchcal{v}}$ acting on basis states as the multiplication one $\hat{\dutchcal{v}}|\dutchcal{v}\rangle=\dutchcal{v}|\dutchcal{v}\rangle$, and is proportional to the oriented volume of the dust ball interior $\hat{\dutchcal{V}} = \alpha_o \hat{\dutchcal{v}}$, where the constant $\alpha_o:=2 \pi \gamma \sqrt{\Delta} \ell_{\mathrm{Pl}}^2$ is in turn expressed via the LQC \emph{area gap} $\Delta$ (which equals $\Delta = 4 \sqrt{3} \pi \gamma \ell_{\mathrm{Pl}}^2$) and Barbero-Immirzi parameter $\gamma$. In further considerations, we take the value of the latter as determined in \cite{Domagala:2004jt, Meissner:2004ju} $\gamma = 0.2375...$.
  \item the holonomy component $\hat{\mathcal{N}}:=\widehat{\exp(i\dutchcal{b}/2)}$ that acts on $|\dutchcal{v}\rangle$ as a unit shift operator: $\hat{\mathcal{N}}|\dutchcal{v}\rangle = |\dutchcal{v}+1\rangle$. The ``sine'' operator stands then for $\sin(\hat{\dutchcal{b}}) = -i/2 (\hat{\mathcal{N}}^2-\hat{\mathcal{N}}^{-2})$. 
\end{enumerate}
The pair $\hat{\dutchcal{v}}, \hat{\mathcal{N}}$ is the quantum counterpart of the (exponentiated for the second element) pair of classical canonical variables $(\dutchcal{v},\dutchcal{b})$ of which Poisson bracket equals $\{\dutchcal{v},\dutchcal{b}\}=- 2/\hbar$, though one has to remember, that in LQC the operator corresponding to $\dutchcal{b}$ does not exist. The basic commutation relation between our elementary operators is $[\hat{\dutchcal{v}},\hat{\mathcal{N}}] = -\hat{\mathcal{N}}$.

In general, the scalar field $\phi$ can be decomposed into an assembly of decoupled harmonic oscillators. As usual in the framework \cite{Ashtekar:2009mb, Parvizi:2021ekr}, we pick a single oscillator with wave-vector $\mathbf{k}$, and write 
\begin{equation} \label{field}
    \hat{H}_\phi = \frac{1}{2}  \alpha_o^{-1} \hat{\dutchcal{v}}^{-1} \otimes \hat{P}_\textbf{k}^2 + \frac{1}{2} k^2 \alpha_o^{1/3} \hat{\dutchcal{v}}^{1/3} \otimes \hat{Q}_\textbf{k}^2 \, ,
\end{equation}
where $\hat{Q_{\mathbf{k}}}, \hat{P_{\mathbf{k}}}$ are respectively position and momentum operators representing single harmonic oscillator with commutation relation $[\hat{Q}_\textbf{k}, \hat{P}_\textbf{k}] = i \hbar$ \cite{Ashtekar:2009mb}.

To ultimately determine the dynamics coming from \eqref{evolution}, we move to so-called \textit{effective dynamics}. Even though solving \eqref{evolution} in the genuine quantum regime is practically possible yet laborious, we can regard the quantum evolution of expectation values of observables as quantum-corrected trajectories living on classical phase space with an effective Hamiltonian. More precisely, all the physical information on the quantum state can be captured by a set of quantum phase space coordinates: expectation values of observables corresponding to basic quantum variables (describing a given system) and the generalized \emph{central Hamburger moments} encoding the quantum corrections in the order-by-order basis\footnote{An example of such objects are variances and correlations of basic variables, being central moments of order $2$.} \cite{Bojowald:2005cw}. The extended coordinates have a well-defined Poisson algebra structure. Furthermore, it is possible to expand each quantum observable as their function. Applying that property to the Hamiltonian it is then possible to write a closed, yet infinite (countable) set of equations of motions capturing the quantum evolution. By introducing a cutoff of this system above certain chosen order one arrives to a finite system of classical equations of motion incorporating quantum corrections up to that order.

The discussed procedure is well understood for the sets of basic observables (quantum counterparts of basic classical variables) forming a Heisenberg algebra. Adapting it for the polymer quantization is a bit more challenging, yet does not pose any significant conceptual challenges. In that context one can ask about a leading order evolution, that is to neglect quantum corrections on the second order and higher. Contrary to a common expectation, this will not bring us back to the classical dynamics, as the modified structure following from polymer quantization will be preserved. In order to see that, we first choose for the geometry degrees of freedom a triad of basic coordinates -- the expectation values $(\dutchcal{v}:=\langle\hat{\dutchcal{v}}\rangle, \, s_{\dutchcal{b}} := \langle(i/2)(\hat{\mathcal{N}}^2-\hat{\mathcal{N}}^{-2})\rangle, \, c_{\dutchcal{b}} := \langle(1/2)(\hat{\mathcal{N}}^2+\hat{\mathcal{N}}^{-2})\rangle)$. For the matter part we in turn select the expectation values of observables corresponding to the original classical canonical variables $(Q_k,P_k)$ with Poisson bracket $\{Q_{\mathbf{k}},P_{\mathbf{k}}\}=1$. Furthermore, we assume that there exist a sufficiently large set of semi-classical states $\Psi_{(\dutchcal{v},s_{\dutchcal{b}}, c_{\dutchcal{b}}, Q_{\mathbf{k}}, P_{\mathbf{k}})}$ peaked on the selected variables. Then, in the zeroth order approximation the effective Hamiltonian 
$H_{\text{eff}} (\dutchcal{v},s_{\dutchcal{b}}, c_{\dutchcal{b}}, Q_{\mathbf{k}}, P_{\mathbf{k}})$ is a zeroth order term in expansion of $\langle \Psi_{(\dutchcal{v},s_{\dutchcal{b}}, c_{\dutchcal{b}}, Q_{\mathbf{k}}, P_{\mathbf{k}})} | \left(\hat{H}_{\text {grav }}+\hat{H}_\phi\right) |  \Psi_{(\dutchcal{v},s_{\dutchcal{b}}, c_{\dutchcal{b}}, Q_{\mathbf{k}}, P_{\mathbf{k}})} \rangle$ becoming
\begin{equation} \label{eff-pre}
    H_{\text{eff}} (\dutchcal{v},s_{\dutchcal{b}}, c_{\dutchcal{b}}, Q_{\mathbf{k}}, P_{\mathbf{k}}) = -\frac{3 \pi G \hbar^2}{2 \alpha} \dutchcal{v} s_{\dutchcal{b}}^2  + \frac{1}{2} \alpha_o^{-1} \dutchcal{v}^{-1} P_{\mathbf{k}}^2 + \frac{1}{2} k^2 \alpha_o^{1/3} \dutchcal{v}^{1/3} \, Q_{\mathbf{k}}^2 \, .
\end{equation}
At this point, we can write a set of equations of motion for the set of chosen $5$ variables. We note, however, that \emph{for as long as we neglected all the higher order quantum corrections} we can reintroduce back the variable $\dutchcal{b}$ such that $\sin(\dutchcal{b}) = s_{\dutchcal{b}}$ and $\cos(\dutchcal{b}) = c_{\dutchcal{b}}$. Upon setting the Poisson structure $\{\dutchcal{v},\dutchcal{b}\}=-2/\hbar$ we reproduce the correct algebra structure of $(\dutchcal{v},s_{\dutchcal{b}},c_{\dutchcal{b}})$, thus bringing the system to the form consistent with that corresponding to the heuristic effective dynamics introduced in \cite{Singh:2005xg}). The effective Hamiltonian will then take the form
\begin{equation} \label{eff}
    H_{\text{eff}} (\dutchcal{v},\dutchcal{b}, Q_{\mathbf{k}}, P_{\mathbf{k}}) = -\frac{3 \pi G \hbar^2}{2 \alpha} \dutchcal{v} \sin^2 (\dutchcal{b})  + \frac{1}{2} \alpha_o^{-1} \dutchcal{v}^{-1} P_{\mathbf{k}}^2 + \frac{1}{2} k^2 \alpha_o^{1/3} \dutchcal{v}^{1/3} \, Q_{\mathbf{k}}^2 \, .
\end{equation}
The Hamilton's equations for it provide us with the set of effective equations of motion, that is
\begin{equation}
\left\{ \begin{aligned} 
    \dot{\dutchcal{v}} &= \frac{3 \pi G \hbar}{ \alpha_o } \dutchcal{v} \sin (2 \dutchcal{b})  \, , \\
    \dot{\dutchcal{b}} &= -\frac{3 \pi G \hbar}{ \alpha_o } \sin^2(\dutchcal{b}) -  \alpha_o^{-1} \hbar^{-1} \dutchcal{v}^{-2} P_{\mathbf{k}}^2  + \frac{1}{3}  \alpha_o^{1/3} \hbar^{-1}  k^2  \dutchcal{v}^{-2/3} Q_{\mathbf{k}}^2 \, ,   \\
    \dot{Q}_{\mathbf{k}} &= \alpha_o^{-1} \dutchcal{v}^{-1} P_{\mathbf{k}} ,  \\    \dot{P}_{\mathbf{k}} &= - k^2 \alpha_o^{1/3} \dutchcal{v}^{1/3} Q_{\mathbf{k}} \, , \\
\end{aligned}
\right.
\end{equation}
where the dot denotes differentiation with respect to $T$. The above system needs to be solved with initial conditions $(\dutchcal{v}_{\text{in}}, \dutchcal{b}_{\text{in}}, Q_{\mathbf{k} \, \text{(in)}}, P_{\mathbf{k} \, \text{(in)}} )$. Throughout this work we set $\dutchcal{b}_{\text{in}} (0) = \pi/2$, which corresponds to setting the ``time origin'' $T=0$ at the \textit{bounce}.

Having defined the description of the dynamics of the collapsing dust ball interior, we are now ready to determine the \textit{rainbow metric}. Suppose that the interior is effectively described by a \textit{classical} FRW line element
\begin{equation} \label{rainbowfrw}
    \tilde{g}_{a b} \mathrm{d} x^a \mathrm{d} x^b=-\tilde{N}(T) \mathrm{d} T^2+\tilde{a}^2(T) \mathrm{d} \mathbf{x}^2 \, .
\end{equation}
One can construct regular QFT on that fixed background and the metric components of (\ref{rainbowfrw}) will be extracted from the proper identification of the scalar field's Hamiltonian ingredients. The corresponding Hamiltonian for a single mode $\textbf{k}$ of the massless scalar field is then given by \cite{Ashtekar:2009mb}
\begin{equation} \label{fieldrainbow}
    \hat{\tilde{H}}_\phi =  \frac{\tilde{N}(T)}{2 \tilde{a}(T)} \left( \hat{P}_{\textbf{k}}^2 + k^2 \tilde{a}(T)^4 \hat{Q}_{\textbf{k}}^2 \right) \, .
\end{equation}
Now we can compare (\ref{fieldrainbow}) to (\ref{field}), in order to relate the elements of the above effective description with variables in which we describe the (effective zeroth order) quantum dynamics
\begin{equation}
    \left\{  
    \begin{aligned}
    \frac{\tilde{N}(T)}{2 \tilde{a}(T)} &= \alpha_o^{-1} \dutchcal{v}^{-1} \, , \\
    \tilde{N}(T) \tilde{a}(T) &= \alpha_o^{1/3} \dutchcal{v}^{1/3} \, . \\
    \end{aligned}
    \right.
\end{equation}
Solving the above yields $\tilde{N}(T) =1 $ and $\tilde{a} = \left( \alpha_o \dutchcal{v} \right)^{1/3}$. When the scalar field is neglected in this system ($Q_{\mathbf{k}} = P_{\mathbf{k}} =0$) an analytic solution for the scale factor is available \cite{Husain:2011tm}  
\begin{equation}
\label{scaleanalytic}
    \tilde{a}(T) \rightarrow a(T) = (\alpha_o \dutchcal{v}(T))^{1/3} = \left( \frac{9 \pi ^2 G^2 \hbar^2 \dutchcal{v}_{\text{in}}}{\alpha_o} T^2 +\dutchcal{v}_{\text{in} } \alpha_o \right)^{1/3} \, , 
\end{equation}
otherwise $a(T)$ has to be determined by numerical methods. Having that, the interior geometry is completely determined. Indeed, one can write the line element with the coordinates $(T,r,\Omega)$ as 
\begin{equation}
\label{frw}
    \mathrm{d}s^2_- = -\mathrm{d}T^2 + a(T)^2 \mathrm{d}r^2  + r^2 a(T)^2 \mathrm{d}\Omega^2 \, ,
\end{equation}
where $\mathrm{d}\Omega = \mathrm{d}\theta^2 + \sin^2 \theta \, \mathrm{d}\varphi^2$. Finally, the mass of the collapsing ball is related with the dust field energy density in the following way
\begin{equation}
\label{mass}
    M = (4 \pi /3 ) \rho R^3 \, ,
\end{equation}
in which $R = r a$ is the physical radius. In the semiclassical approximation taken, the dust energy density takes the form $\rho = p_T /a^3$, where $p_T = - H_{\text{eff}}$ \cite{Parvizi:2021ekr}.

\subsection{Exterior metric and junction conditions} \label{secexterior}

In the considered model, the exterior of the collapsing ball is a spherically symmetric asymptotically flat\footnote{This requirement, while incorporated for a physical reason does not affect the following considerations. The asymptotic flatness has not been imposed as a condition in any step of the calculations.} (region of) spacetime that is either vacuum (in absence of the scalar field) or contains a residual scalar field corresponding to the particular single mode selected in constructing the rainbow metric scenario. In classical GR the process of dust ball collapse has been studied extensively (see for example Chapter 3.8 of \cite{Poisson:2009pwt}). In particular, it is a well-established result, that starting from the requirements of spherical symmetry and staticity one ends up in the class of metrics that can be expressed in Eddington-Finkelstein coordinates. We carry this insight as a well-motivated assumption to this (quantum geometry modified) scenario\footnote{This expectation has been in fact confirmed in a followup \cite{Lewandowski:2022zce} of the preliminary presentation \cite{Bobula2022} of our results.}, essentially following (though in more precise manner) the analysis of \cite{Parvizi:2021ekr}.

To be more precise, in what follows we assume that there exists an exterior region described by the following form of the metric written in either ingoing Eddington-Finkelstein coordinates $(v, X, \Omega)$ 
\begin{equation} \label{finkelin}
    \mathrm{d}s_+^2 = -F(X,v) \mathrm{d}v^2 + 2 \mathrm{d}v \mathrm{d}X + X^2 \mathrm{d} \Omega^{2} \, ,
\end{equation}
or the outgoing ones\footnote{Adding this class of metric is necessary due to dynamic bounce of the dust ball interior.}, with $(u, X,\Omega)$
\begin{equation} \label{finkelout}
    \mathrm{d}s_+^2 = -F(X,u) \mathrm{d}u^2 - 2 \mathrm{d}u \mathrm{d}X + X^2 \mathrm{d} \Omega^{2} \, .
\end{equation}
The regions described by each of the above systems are apriori distinct, yet in principle, an overlap between them may (and in fact, as we will show further in, does) exist.

We further assume, that in both the incoming and outgoing region the function $F$ does depend on the coordinate $X$ only, Physically this is a requirement of the existence of a (timelike) Killing field in the exterior of the dust ball. This is a strong assumption, that will significantly affect the global spacetime structure of the system and its consequences will be discussed in the conclusions.

In the coordinate system used to characterize the interior the surface $\Sigma$ of the collapsing ball is parameterized as $ x^\alpha = (T, r=r_b = const.$), which is also the trajectory of the comoving inertial observer (the dust particle at the ball surface). Alternatively, in the region where the exterior is described by \eqref{finkelin} it can be written in the exterior coordinates as  $ x^\alpha = (v = V(T), X = R(T))$\footnote{Although we have already named the physical radius in \eqref{mass} as $R$, we also use this letter to name the $X = R(T)$ on $\Sigma$. We will show below that these quantities are in fact identical.}. Now let $y^a = (T, \theta, \varphi) $ be a coordinate system on $\Sigma$. The metric induced respectively from the interior/exterior one takes there the form
\begin{equation}
    \mathrm{d}s_-^2 \Big|_\Sigma := h_{ab}^- \mathrm{d} y^a \mathrm{d} y^b= - \mathrm{d} T^{2} + r_b^2 a^2 \mathrm{d} \Omega^2 \, ,
\end{equation}
and
\begin{equation}
    \mathrm{d}s_+^2 \Big|_\Sigma := h_{ab}^+ \mathrm{d} y^a \mathrm{d} y^b = -( F \dot{V}^2 - 2\dot{V} \dot{R}) \mathrm{d} T^2 + R^2 \mathrm{d} \Omega^2 \, ,
\end{equation}
where the dot denotes differentiation with respect to $T$. The four-velocity of an observer comoving with the surface is $l^\alpha = \partial x^\alpha / \partial T$. Let $n^\alpha$ be a normal to $\Sigma$, we impose $n^\alpha n_\alpha = 1$, $n^\alpha l_\alpha = 0$ (we then choose $n^\alpha$ so that it points outside the collapsing ball). Then its components are $n^-_T = 0$, $n_r^- = a$ or 
\begin{equation}
    n_{v}^+ = - \frac{\dot{R} / \dot{V} }{\sqrt{F-2\dot{R}/\dot{V}}} \, ,
\end{equation}
\begin{equation}
    n_X^+ = \frac{1}{\sqrt{F - 2 \dot{R}/\dot{V}}}\, .
\end{equation}
Now we can write junction conditions:
\begin{enumerate}
  \item The continuity of metric time-time component $h_{TT}^-=h_{TT}^+$ implies
\begin{equation} \label{one}
  F \dot{V}^2 - 2\dot{V} \dot{R} =1 \, ,
   \end{equation}
\item whereas from $h_{\theta\theta}^-=h_{\theta\theta}^+$ we have
  \begin{equation} \label{two}
         R = r_b a \, .
   \end{equation}
\item The requirement of the continuity of the extrinsic curvatures $K_{ab}^+ = K_{ab}^-$, where $K_{ab} := n_{\alpha;\beta} \frac{\partial x^\alpha}{\partial y^{a}}\frac{\partial x^\beta}{\partial y^{b}}$ \cite{Poisson:2009pwt}  gives 
\begin{equation} \label{three}
    F n^+_X + n_v^+ = 1 \, .
\end{equation}
as a consequence of $K^-_{\theta \theta} = K^+_{\theta \theta}$ and 
\item similarly, 
\begin{equation} \label{four}
n_{\alpha ; \beta } l^\alpha l^\beta = 0 \, ,
\end{equation}
follows from $K^-_{T T} = K^+_{T T} $.
\end{enumerate}

Following \cite{Parvizi:2021ekr}, we would like to encode the physical volume of collapsing ball as $\dutchcal{V} = \alpha_o \dutchcal{v} = a^3$ which follows from the cosmological origin of the implemented treatment. This choice via \eqref{two} (from which follows, that $\dutchcal{V} = (4/3) \pi R^3$) fixes the ball comoving coordinate radius as $r_b = (4 \pi / 3)^{-1 / 3}$. 

The above system of junction conditions is soluble. Equations (\ref{one},\ref{two},\ref{three}) combined with an additional requirement, that the exterior metric is of the Lorentzian signature: $F-2\dot{V}/\dot{R} > 0$, $F<1$ yield
\begin{equation} \label{fsigma}
    F = 1 - \dot{R}^2 \, ,
\end{equation}
on the surface $\Sigma$. That relation will be crucial in the following analysis. Due to assumed symmetries of the line element (\ref{finkelin}), we can invert the relation (\ref{two}) to find $T^2(X)$ and subsequently write $F(T^2(X))$. For the case of pure dust ball collapse, where the scale factor is given by \eqref{scaleanalytic} we have
\begin{equation} \label{t2}
    T^2 (X) = \frac{\alpha_o \left(X^3-\alpha_o r_{b}^3
   \dutchcal{v}_{\text{in}}\right)}{9 \pi ^2 G^2 \hbar^2 r_{b}^3 \dutchcal{v}_{\text{in}}} \, .
\end{equation}
Thus
\begin{equation}
    F(X) = 1 - \dot{R}^2 (T^2(X)) =1 -\frac{4 \pi ^2 G^2 \hbar^2 r_b^3 \dutchcal{v}_{\text{in}}}{\alpha_o X}+\frac{4 \pi ^2 G^2 \hbar^2 r_b^6
   \dutchcal{v}_{\text{in}}^2}{X^4} \, .
\end{equation}
According to (\ref{mass}), we can identify the mass of the dust ball to be $M=2 \pi^2 G \hbar^2 r_b^3 \dutchcal{v}_{\text{in}}/\alpha_o$. Then, the above formula takes a more familiar form
\begin{equation} \label{FXanal}
F(X) = 1 - \frac{2 G M}{X} + \frac{2 G M \alpha_0 r_b^3 \dutchcal{v}_{\text{in}}}{ X^4 }\, .
\end{equation}
The metric component (\ref{FXanal}) has two roots $X_-$ and $X_+$ (where we choose the order $X_+ > X_-$) for masses 
\begin{equation}\label{eq:Mext}
  M> M_\text{ext}:= 16 \alpha_o r_b^3/(9 \sqrt{3} G^2 \hbar) \approx 0.83 \, m_{\text{Pl}} 
\end{equation}
(equivalently for  $\dutchcal{v}_{\text{in}} > 8 \alpha_o^2/(9 \sqrt{3} \pi ^2 G^3 \hbar^3)$), exactly one root for $M=M_\text{ext}$, and no roots for $M<M_\text{ext}$. Throughout the rest of this work, we restrict the discussion to $M\gg M_\text{ext}$, since only then the semiclassical approximation used to determine the effective interior metric is sufficiently accurate. For low masses, especially those of the order of $m_{{\rm Pl}}$ the quantum origin of the metric makes the presented treatment unreliable. This issue will be discussed in more detail in the conclusions.

When the scalar field is included, the algorithm specified above is still applicable, however now (due to lack of a closed formula for $a(T)$) the function $F(X)$ can be determined only numerically.

At this point, $R(T)$ can be easily extracted since the explicit form of the scale factor is known, so it remains to solve for $V(T)$ to completely determine the trajectory of the dust ball. Equations (\ref{one}, \ref{fsigma})\footnote{We can rewrite \eqref{fsigma} as $\dot{R} = \text{sgn}(T) \sqrt{1-F}$ since $\dot{R} > 0$ for $T>0$, and  $\dot{R} < 0$ for $T<0$. That is because of the time reflection symmetry $T \rightarrow -T$ of \eqref{scaleanalytic} (note that $R = r_b a$). The same situation is for the case when one incorporates (a single mode of) scalar field (scale factor is given numerically).  }  combined together give

\begin{equation} \label{Vdot}
    \dot{V}(T) = \frac{\text{sgn}(T) \sqrt{1 -F(T)} +1}{F(T)} \, .
\end{equation}
 We employ the following strategy for solving the above. First, we notice that the function $j(T):=\dot{V}(T)(T-T_1)(T-T_2)$ is regular for all times $T$, where $T_1 := T(X_-) >0$ and $T_2:=T(X_+) >0$ accordingly to \eqref{t2}. This property allows us to define
\begin{equation}
    \dot{\Tilde{V}}(T) := \dot{V}(T) - \frac{C_1}{(T - T_1)} - \frac{C_2}{(T-T_2)}\, ,
\end{equation}
 with constants $C_1 = j(T_1)/(T_1 - T_2) $ and $C_2 = -j(T_2)/(T_1 - T_2)$ which is also regular.  Finally, $V(T)$ can be determined via integrating out the regular and singular terms of $\dot{V}(T)$  
 \begin{equation}
     V(T) - V(0) = \int^T_0 \dot{\Tilde{V}}(T) \, \mathrm{d}T + C_1 \log|T-T_1| + C_2 \log |T-T_2| - C_1 \log|T_1| - C_2 \log|T_2| \, .
 \end{equation}
 
 The similar construction can be performed for the outgoing line element \eqref{finkelout} by repeating the above steps. In that case, we extract an equation
 \begin{equation}
    \label{Udor}
\dot{U}(T) = \frac{-\text{sgn}(T) \sqrt{1 -F(T)} +1}{F(T)}\, ,
\end{equation}
which can be analogously solved -- the qualitative difference is the fact that the singular points are time reflected $T \rightarrow -T$. We emphasize that $V(T), U(V)$ deliver coordinate values at the $\Sigma$, and also label ingoing/outgoing surfaces (geodesics) respectively -- the more detailed analysis will be presented at the end of this section.

\begin{figure}[h] 
   
    \centering
    \includegraphics[width=11cm]{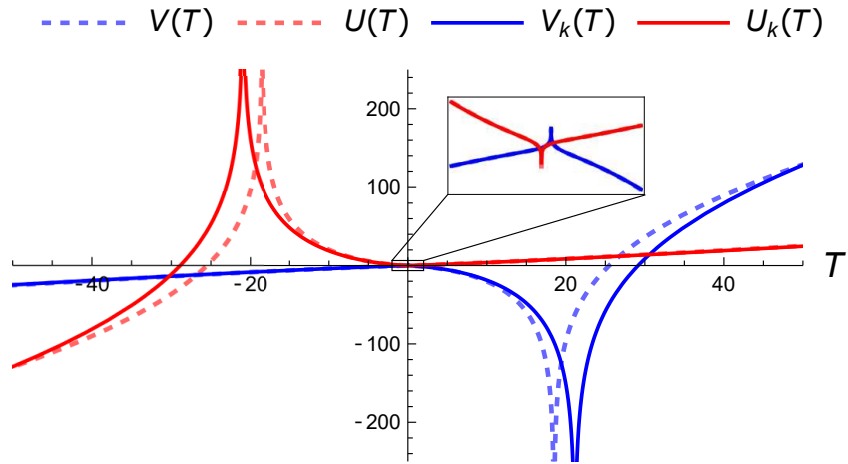}
    \caption{ Time coordinate trajectories ($V(T),V_k(T) $) and ($U(T),U_k(T)$) of the surface of the collapsing dust ball. Initial conditions are $\dutchcal{v}_{\text{in}} = 10$, $\dutchcal{b}_{\text{in}} = \pi/2$, $Q_{\mathbf{k}, \text{(in)}} = P_{\mathbf{k}, \text{(in)}} = 0.3$, $V(0)=V_k(0)=0$, $U(0)=U_k(0)=0$ and also $k =G = c = \hbar =1 $, $\gamma = 0.23\dots$ . Pairs of singular points are visible, they reflect the existence of horizons.} 
     \label{UV}
\end{figure}

 In Figure \ref{UV}, we present the solutions $V(T), U(T)$ for pure dust ball collapse (the scale factor (\ref{scaleanalytic}) is incorporated) and for the dust perturbed with a single mode of the scalar field $V_{k}(T), U_k(T)$ (the scale factor is given numerically, however, \eqref{Vdot}, \eqref{Udor} still hold).

\begin{figure}[h] 
    \centering
    \includegraphics[width=4cm]{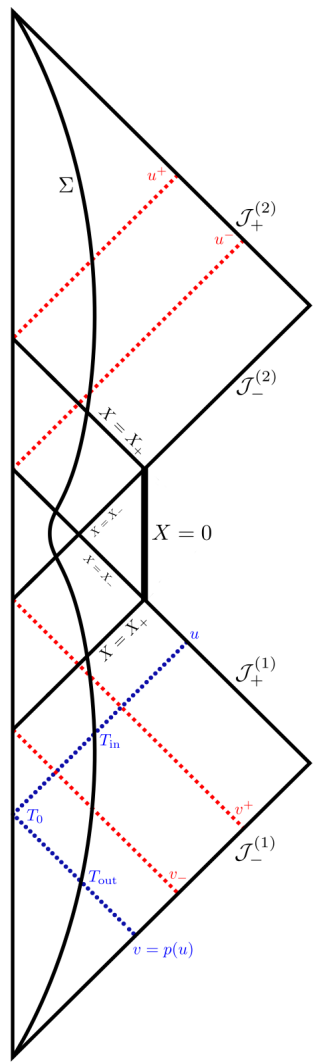}
    \caption{Conformal diagram for the resulting exterior space-time. The ball of matter with the surface $\Sigma$ collapses and then expands in a new \textit{universe} after the \textit{bounce} at $T=0$. Inner $X=X_-$ and outer $X=X+ $ horizons are formed. Timelike singularity is present at $X=0$.}
     \label{conformal}
\end{figure}

Now, we wish to determine the global causal structure of the resulting spacetime. To achieve so, it is convenient to introduce a metric in double null coordinates. Line elements (\ref{finkelin}, \ref{finkelout}) possess one piece of the ingoing and the outgoing null coordinate respectively. Notice that the constant values of those at $\Sigma$ label corresponding ingoing/outgoing null surfaces in the exterior. Since we have solved for $V(T), U(T)$ for all times of $T$, we have obtained ``foliation'' of the exterior with either ingoing or outgoing null surfaces. Hence, there exist regions of the exterior where (\ref{finkelin}, \ref{finkelout}) overlap. The consistency of the description demands that, for example, the null condition $\mathrm{d} u = 0$ (coming from \ref{finkelout}) is the same as the condition $\mathrm{d}v - \frac{2}{F(X)} \mathrm{d} X =0$ (coming from \ref{finkelin}). This is equivalent to the following relation
\begin{equation} \label{doublenull}
\mathrm{d} v - \mathrm{d}u = \frac{2}{F(X)} \mathrm{d} X \, .
\end{equation}
Plugging the above to either (\ref{finkelin}) or (\ref{finkelout}) makes the exterior to be described with $(u,v, \Omega)$ and
\begin{equation}
\label{null}
    \mathrm{d}s^2_+ = - F(X) \mathrm{d}u\mathrm{d}v + X^2 \mathrm{d}\Omega^2 \, .
\end{equation}
Next, we want to find out what are the boundaries of the exterior. We study the behaviour of an affine parameter on null geodesics (surfaces).  Consider $u=constant$, radial null geodesic, and let $v$ be the parameter. The parametrization can be represented as $(u(v) = constant, v, 0 ,0)$, so the tangent vector is $m^\alpha$ = (0, 1, 0, 0). The corresponding four-acceleration satisfies $m^\alpha_{;\beta}m^\beta = \frac{1}{2} \frac{\mathrm{d}F(X) }{\mathrm{d} X} m^{\alpha}$. It means that $v$ is not affine. However, we can extract the affine parameter, say $\lambda$, specifically from $\frac{\mathrm{d}^2\lambda(v)}{\mathrm{d}v^2} = \frac{1}{2}\frac{\mathrm{d}F}{\mathrm{d}X} \frac{\mathrm{d} \lambda(v)}{\mathrm{d} v}$ (an arbitrary parameter always can be related to the affine one \cite{Poisson:2009pwt}). When we implement the condition (\ref{doublenull}), particularly 
\begin{equation} \label{nullcondition}
\frac{\mathrm{d}X}{\mathrm{d}v} = \frac{1}{2} F(X) \, ,
\end{equation}
we arrive at 
\begin{equation}
  \frac{\mathrm{d}^2\lambda(X)}{\mathrm{d}X^2} =0 \, ,     
\end{equation}
so $X$ is the affine parameter (that is not true when $F(X)=0$). Next, the strategy is to analyze how $X$ varies on the whole family of $u=constant$ geodesics. Suppose we start moving along the null geodesic towards the future from a point at $\Sigma$. To see how $X$ changes in time $v$, it is necessary to solve (\ref{nullcondition}) with a proper initial condition. Given a point $T=constant$ at $\Sigma$, we can read off the initial $v = V(T)$ from Figure \ref{UV}. We conclude that for starting points $T \in (-\infty, -T(X_+)) \cup (-T(X_-), \infty)$, where $T(X_-)$, $T(X_+)$ -- singular points of Figure \ref{UV} -- are taken to be positive according to (\ref{t2}), the affine parameter $X$ increases and reaches $X \rightarrow \infty$ when we move along the geodesic towards the future. Surprisingly, for starting points $T \in (-T(X_+),-T(X_-))$, $X$ decreases and reaches $X\rightarrow 0$. At that limit, the Kretschmann scalar diverges, which reveals the existence of coordinate-independent singularity, particularly the timelike one. The analogous steps are done for the $v=constant$ family of geodesics, and the conclusions concerning the global picture stay the same. Also, we notice that surfaces of $X=X_+$ and $X=X_-$ are null -- the right-hand side of \eqref{nullcondition} vanishes. They represent black/white hole event and inner horizons respectively. We summarize the above construction in Figure \ref{conformal}.

\section{Entanglement entropy}

With the global spacetime structure of the model established, we can now turn our attention to the main topic of this article -- the problem of information loss. In its essence, it is a loss of unitarity in the evolution of quantum fields over a spacetime containing an evaporating black hole. In particular, the pure state of a matter (for example, a scalar field) incoming from past null infinity becomes a mixed state at the future null one. In an attempt to measure this process several notions of entropy associated with the field have been introduced, in particular some definitions of the \emph{entanglement entropy}. In our work, we will use the
\textit{renormalized entanglement production}. The calculation will reveal qualitative properties relevant in the context of the paradox. In particular, we want to verify whether we can obtain characteristic up-then-down behaviour of the (paradox-free) Page curve \cite{ Page:1993wv, Holzhey:1994we, Bianchi:2014qua, Bianchi:2014bma}. The entropy essentially quantifies how much entanglement is produced by a test massless scalar field when evolved from the past null conformal infinity (further denoted as $\mathcal{J}_-$) to the future null one (denoted as $\mathcal{J}_+$) in asymptotically flat spacetime. Specifically, it measures how entangled the field vacuum is (when evolved from $\mathcal{J}_-$ to $\mathcal{J}_+$) in a subregion of $\mathcal{J}_+$ to its complement (in principle full $\mathcal{J}^+$ corresponds to a pure state). 
The actual entropy calculation procedure employs
the so-called \textit{ray-tracing map}, where one sends outgoing light rays terminating at $\mathcal{J}_+$ all the way back to their origin at $\mathcal{J}_-$ and measures their ``compression'' with respect to affine null parameters normalized using the asymptotic structure. More precisely, let $u$ and $v$ be affine parameters at $\mathcal{J}_+$ and $\mathcal{J}_-$ respectively. Light rays marked by $u$ are travelled back in time along radial null geodesics (geometric optics approximation) so that the mapping $p(u) = v$ relates the parameters associated with each ray reaching $\mathcal{J}_-$. The entanglement entropy production of the region $[-\infty, u]$ can then be evaluated via a generalization of the Larsen-Holzhey-Wilczek entropy \cite{Holzhey:1994we} for spherically symmetric and asymptotically flat space-times \cite{Bianchi:2014qua}
\begin{equation}
\label{entropy}
    S(u) := \lim_{u_0 \rightarrow -\infty}  S(u, u_0) = \lim_{u_0 \rightarrow -\infty} \frac{1}{12} \log \frac{(p(u) - p(u_0))^2}{\dot{p}(u) \dot{p}(u_0) (u-u_0)^2} = -\frac{1}{12} \log \dot{p}(u) \, .
\end{equation}
Here, the dot denotes differentiation with respect to $u$. The last equality is satisfied in cases when $\underset{u\to -\infty}{\lim} \dot{p}(u) = 1 $ holds (which due to the complicated structure of the spacetime in our model is not a given here). The finiteness of $S(u)$ at all times of $u$ is a necessary condition for the unitarity of the evolution \cite{Bianchi:2014bma}. If $S(u)$ is negative it means that the state is less correlated than the Minkowski vacuum \cite{Bianchi:2014bma}. 
 
% In our model of gravitational collapse, we calculate the above entropy. 

For our model, with its particular global spacetime structure (see Figure \ref{conformal}) both the future and past null infinity consist of two disconnected portions each. The ray-tracing then relates the first connected portion $\mathcal{J}_+^{(1)}$ to only a part of $\mathcal{J}_-^{(1)}$, whereas the rays terminating at $\mathcal{J}_+^{(2)}$ originate at both $\mathcal{J}_-^{(1)}$ and $\mathcal{J}_-^{(2)}$. 

In order to perform the actual calculation, we first note, that with the form of the line element \eqref{null} we have $\underset{u\to -\infty}{\lim} F(X(u,v)) = \underset{v\to \infty}{\lim} F(X(u,v)) = 1 $, thus the double null coordinates chosen earlier $(u,v)$ \emph{already} are the desired parameters asymptotically affine at null infinities. 
The specific $p(u)$ evaluation algorithm is the following. We first recall, that inside the dust ball, there exists a distinguished global (within the ball) time $T$ -- proper time of dust particles constant on the homogeneity surfaces. Then to each ray, we can associate the time $T_{\rm in}$ of entering and $T_{\rm out}$ of exiting the ball -- see Figure \ref{conformal}. Since the chosen coordinate $u$, $v$ is constant on, respectively, the incoming and outgoing portion of the ray, the values of these coordinates are, respectively $U(T_{\rm in})$ and $V(T_{\rm out})$ satisfying \eqref{Udor} and \eqref{Vdot}. The function $p(u)$ appearing in \eqref{entropy} can then be written as
\begin{equation}\label{eq:p-function}
  p(u) = V(T_{\rm out}(T_{\rm in}(u))) \, ,
\end{equation}
Note, that the function $U(T_{\text{in}})$ is only piecewise invertible, thus $T_{\rm in}(u)$ is not globally well defined, which will have crucial consequences for the final result.

In order to find the last missing relation $T_{\rm out}(T_{\rm in})$ we note, that the difference $T_{\rm out}-T_{\rm in}$ is simply the time of traversing the dust ball, which can be easily found by integrating the null condition coming from (\ref{frw})
\begin{equation}
\label{nullfrw}
    \frac{\mathrm{d}T}{\mathrm{d}r } = \pm \, a(T) \, .
\end{equation}
Each ray enters the collapsing dust ball at a point $(T_\text{in}, r_b)$, transits the origin of the radial coordinate $(T_0, r=0)$, and finally, reemerges at the point $(T_\text{out}, r_b)$. The relation $T_{\rm out}(T_{\rm in})$ can be found as follows. Given $T_{\text{in}}$, we can solve for $T_{0}$ from $\int^{T_0}_{T_{\text{in}}} 1/a(T) \, \mathrm{d}T  = \int^{0}_{r_b} \mathrm{d} r$, consecutively for $T_\text{out}$ from $\int^{T_\text{out}}_{T_0} 1/a(T) \, \mathrm{d}T  = -\int^{r_b}_{0} \mathrm{d} r$. The closed form of each of the above integrals with the scale factor \eqref{scaleanalytic} incorporated can be performed analytically, however, the solution is not practically useful for deriving the closed form of desired $T_{\rm out}(T_{\rm in})$ -- the integration outcomes include hypergeometric functions. Hence, we solve for $T_{\rm out}(T_{\rm in})$ numerically in both cases -- the pure dust ball collapse and the collapse perturbed with the single mode of the scalar field.

\begin{figure}[h] 
    \centering
    \includegraphics[width=11cm]{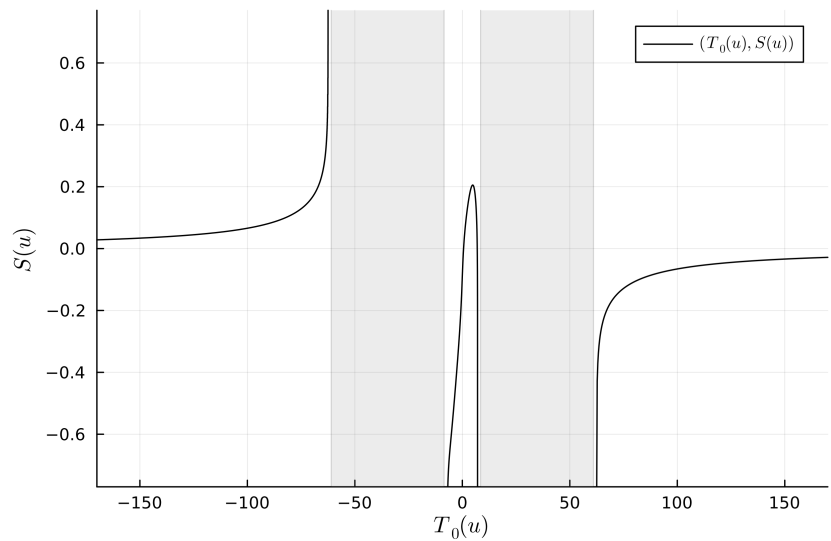}
    \caption{Parametric plot of $(T_0(u), S(u))$ for the pure dust ball collapse. $T_0(u)$ is taken at $r=0$ -- see Figure \ref{conformal}. Solutions for $U(T), V(T)$ from Figure \ref{UV} were utilized. All possible null infinities were related. Three distinct shapes are visible for light rays: a) semiclassical past region, b) deep region, and c) semiclassical future region.  Greyed-out windows correspond to the existence of the timelike singularity. $u_1 = 4$.}
     \label{rysS}
\end{figure}

Even though the interior metric as well as $F(X)$ have simple analytic form, the relation $p(u)$ could not be found analytically. Instead, the relation itself and the entropy have been determined via numerical methods. For that, libraries of \textsc{Julia} programming language have been used. An example of the result is presented in Fig.~\ref{rysS}. For the presentation convenience, the entropy function is parametrized by the time $T_o$ at which a given ray crosses the coordinate system origin $r=0$. We note several features, some of which are quite concerning.

First of all, the proposed entropy evaluation could not be performed globally, as there exist regions of (both future and past) null infinities where the rays either originate or terminate at the singularity and the presented method (strongly relying on having right asymptotic behaviour) cannot be applied. They are represented by two greyed-out windows on fig.~\ref{rysS}.

In the regions, for which the considered null geodesics are complete we can distinguish three domains:
\begin{enumerate}
  \item Semiclassical past region, where the rays labelled with $v \in (-\infty, v^-)$ originate at (a portion of) $\mathcal{J}_-^{(1)}$, cross the collapsing dust ball while never cross the black hole outer horizon and terminate at $\mathcal{J}_+^{(1)}$, where they are labelled by $u \in (-\infty, \infty)$. The rightmost formula in \eqref{entropy} can be applied directly. We observe the entanglement entropy growing and finally reaching infinity as the rays approach the horizon. Indeed, we see that in the vicinity of $u \rightarrow \infty$ (or $T_\text{in}(u) \rightarrow -T_2$) 
  \begin{equation}
    \dot{p}(u) =  \frac{\partial V}{\partial T_\text{out}} \frac{\partial T_{\text{out}}}{\partial T_\text{in}} \left(\frac{\partial U}{\partial T_\text{in}  } \right)^{-1} \propto -\left(T_\text{in}(u)+T_2 \right) \, ,
  \end{equation}
  from which follows $S(u) = -\frac{1}{12}\log \dot{p}(u) \rightarrow \infty$, as $u\to\infty$. 

  \item The deep region, where the rays labelled with $v \in (v^+, \infty)$ originate at $\mathcal{J}_-^{(1)}$, cross both the inner and outer horizon of a black hole as well as (subsequently) the white hole ones, then terminate at $\mathcal{J}_-^{(2)}$. They are labeled by $u \in (-\infty, u^-) $, where $u^-$ denotes the latest ray terminating at $\mathcal{J}_-^{(2)}$ -- see Figure \ref{conformal}. These rays are particularly relevant for ``probing'' the modified quantum gravity, as they cross the high (near Planckian) energy density region, where the quantum geometry effects become dominant. However, the rightmost equality of \eqref{entropy} does not hold. In the vicinity of $u \rightarrow -\infty$ (or $T_\text{in}(u) \rightarrow -T_1$) we have 
  \begin{equation}
    \dot{p}(u) = \frac{\partial V}{\partial T_\text{out}} \frac{\partial T_{\text{out}}}{\partial T_\text{in}} \left(\frac{\partial U}{\partial T_\text{in}  } \right)^{-1} \propto T_\text{in}(u)+T_1 \, ,
\end{equation} 
thus in the desired limit $\dot{p}(u) \rightarrow 0$, which violates the necessary condition for the applicability of the discussed equality, which is $\dot{p}(u) \rightarrow 1$). Furthermore, the limit in the middle expression of \eqref{entropy} cannot be taken directly in its present form. This happens, because in the vicinity of $u_0 \rightarrow -\infty$ (or $T_\text{in}(u_0) \rightarrow -T_1$) 
\begin{equation}
  (p(u_0))^2/(( \dot{p}(u_0) \, u_0^2 )) \propto  1/(\left(\log|T_\text{in}(u_0)+T_1|\right)^2 (T_\text{in}(u_0)+T_1)) \, .
\end{equation}
In consequence, $S(u) = \underset{u_0 \to -\infty}{\lim} S(u, u_0) =\infty$ -- the formula returns infinity for all $u \in (-\infty, u^-) $. One can ask however if this formula can be modified in a way that preserves the physical meaning of the entropy and gives us finite quantities. In order to do so, we recall that \eqref{entropy} is the limit of the regular function $S(u, u_0)$, where $u_0 \rightarrow -\infty$ constitutes ``reference point'' from which the entropy is calculated (produced). The possible construction \emph{generalizing} the above may be about the choice of that point, provided that the modification is unambiguous. With the reference point being changed, the redefinition will be accurate up to an additive constant -- one loses the possibility of determining the beginning point in which the entropy takes the zero value. Note that entropy difference (according to the formula \eqref{entropy}) with respect to two distinct points, $u$ and $u_1$, is given by  
\begin{equation} \label{newentropy}
     \Delta S (u, u_1) := \lim_{u_0 \rightarrow -\infty } \left( S(u,u_0)  - S(u_1, u_0)\right) = \frac{1}{12} \log  \frac{ (p(u) - v^+)^2 \, \dot{p}(u_1) }{(p(u_1) - v^+)^2 \, \dot{p}(u)} \, ,
\end{equation}
where we identified $ \underset{u_0 \to -\infty}{\lim} p(u_0)  = v^+$. 
In order to extract the relevant properties of the above quantity, we numerically check the population of the solutions (for various $M$) as well as we analytically probe its asymptotic behaviour, as $u$ approaches the boundaries of the deep region from its interior. In the vicinity of $u \rightarrow u^-$ (or $T_{\text{out}}(T_\text{in}(u)) \rightarrow T_1$) we have 
\begin{equation}
  p(u)^2/\dot{p}(u) \propto -\left( \log|T_{\text{out}}(T_\text{in}(u))-T_1|  \right)^2 (T_{\text{out}}(T_\text{in}(u))-T_1)
  \quad \Rightarrow\quad 
  \underset{u \to u^-}{\lim} \Delta S (u, u_1) = - \infty  \, .
\end{equation}
Also, in the vicinity of $u \rightarrow -\infty$ we have
\begin{equation}
\Delta S (u, u_1) \propto \log  \frac{ (p(u) - v^+)^2}{\dot{p}(u)} \, .
\end{equation}
Due to the complicated behavior, we inspect the expression on the right hand side of the above formula numerically. The result is $\underset{u \to -\infty}{\lim} \Delta S (u, u_1) = - \infty$. With the established limits, we conclude that the formula \eqref{newentropy} (actually being the entropy difference accordingly to \eqref{entropy}) characterizes the unambiguous (up to the additive constant) entanglement entropy production in the deep region. We depict the result in Figure \ref{rysS} -- the entropy curve rapidly grows from $-\infty$, reaches a finite maximum, and then again rapidly goes back to $-\infty$.
  \item Semiclassical future region, where the rays labelled with $v \in (-\infty, \infty) $ originate at $\mathcal{J}_-^{(2)}$, again traverse the dust ball in its expanding phase (after the bounce), never crossing any of the black/white hole horizons and terminate at $\mathcal{J}_-^{(2)}$. The light rays of consideration are labeled by $u \in (u^+, \infty)$, where $u^+$ denotes the earliest ray terminating at $\mathcal{J}_{+}^{(2)}$. The limit $\underset{u \to -\infty}{\lim} \dot{p}(u)$ cannot be even considered since there are no rays $u<u^+$ (as they would need to originate at the singularity) -- the rightmost equality of \eqref{entropy} does not hold. Moreover, $\underset{u_0 \to u^+}{\lim} S(u, u_0) = -\infty$, since in the vicinity of $u_0 \rightarrow u^+$ (or $T_{\text{out}}(T_\text{in}(u_0)) \rightarrow T_2$)
  \begin{equation} 
    (p(u_0))^2/( (\dot{p}(u_0) \, u_0^2 )) \propto (\log|T_{\text{out}}(T_\text{in}(u_0))-T_2|)^2 \, (T_{\text{out}}(T_\text{in}(u_0))-T_2) \, .
  \end{equation}
  However, the following limit holds $\underset{u_0 \to \infty}{\lim} \dot{p}(u) =1$, and an inverted limit, namely $u_0 \to \infty$, in \eqref{entropy} (also) implies the rightmost equality. Indeed, we argue that for the future region, the formula $ S(u) = -\frac{1}{12} \log \dot{p}(u)$ unambiguously corresponds to the entropy production, however, the actual entropy is now defined with respect to the ``reference point'' (see the discussion for the deep region) taken to be at $u_0 \rightarrow \infty$. We note that $\dot{p}(u) \propto 1/(T_\text{out}( T_\text{in}(u)) - T_2) $ in the vicinity of  $u_0 \rightarrow u^+$ (or $T_\text{out}(T_\text{in} (u_0)) \rightarrow T_2$), so $S(u) \rightarrow -\infty$. That fact implies that  $S(u)$ is calculated with accuracy up to an additive constant. Ultimately, the entropy grows from $-\infty$, and the growth rate slows as the time passes from the point corresponding to the white-hole horizon -- see Figure \ref{rysS}.
\end{enumerate}

\begin{figure}[h] 
    \centering
    \includegraphics[width=8cm]{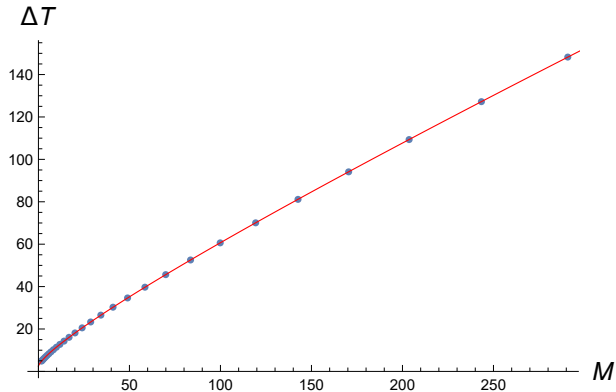}
    \caption{$\Delta T (M)$ for the pure dust ball collapse. Numerical results and fitted function $ \Delta T (M) = 1.372(20) M^{0.6331(52)}+0.3305(28) M+2.318(38)$.}
     \label{timeandmass}
\end{figure}

\begin{figure}[h] 
    \centering
    \includegraphics[width=11cm]{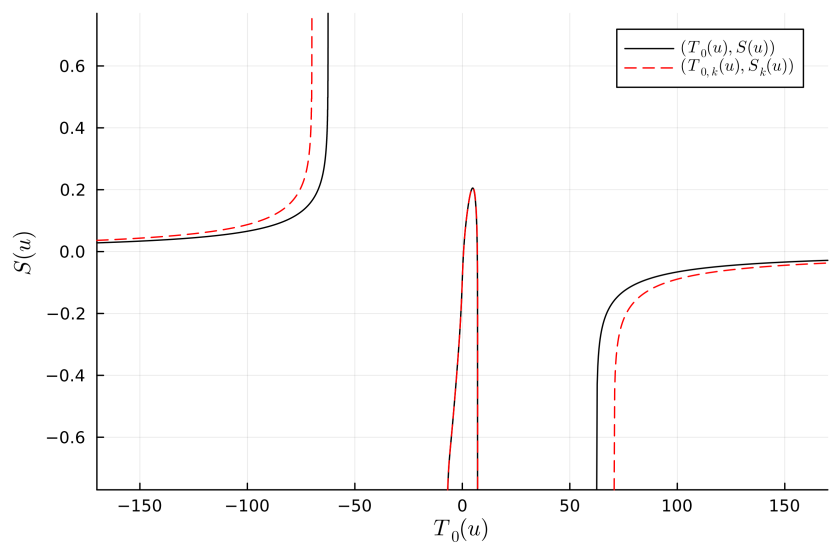}
    \caption{A comparison of $(T_0(u), S(u))$ (pure dustball collapse) and $(T_0(u), S_k(u))$ (dustball perturbed by scalar field). Analogously as in Figure \ref{rysS}, $U_k(T), V_k(T)$ were utilized to calculate $S_k(T)$. $u_1 = 4$.}
     \label{rysSks}
\end{figure}

One particularly interesting question is the one about the length of the epoch, during which the rays interact with the quantum-geometry-modified region. In cosmology, the epoch, when the geometry discreteness effects dominate the dynamics is extremely short -- well below Planck second. Since the model of the interior is that of LQC, the length of the epoch when the dust energy density is comparable to the Planck one will be of the same order. However, the time (as measured by the observer at the origin $r=0$) during which the rays interacting with the quantum-modified regime are passing through the origin may be much longer. In order to determine that we
define the epoch length as the difference in crossing the origin via the earliest and latest ray of the distinguished epoch (spacetime region) $\Delta T = T_{r=0}^{+} - T_{r=0}^{-}$, where $ T_{r=0}^{+}$ is the time at $r=0$ when we traversed along the null geodesic to the future from $T_{r=r_b}(X_-) >0$ up to $r=0$. Analogously, $T_{r=0}^{-}$ is the time when we traversed to the past from $T_{r=r_b}(X_-) <0$ to $r=0$. $\Delta T$ is also the time of entanglement entropy production in the deep region -- see Figures \ref{rysS}, \ref{rysSks}. We numerically calculate $\Delta T$ for different masses $M$ -- see Figure \ref{timeandmass}. For the mass of the black hole in the centre of our galaxy (Sagittarius A*), we have $\Delta T = 1.25(11) \times 10^{44}  \, t_\text{Pl}$, which is roughly a few seconds. For the solar mass we have $\Delta T = 3.020(86) \times 10^{37} \, t_\text{Pl}$, which is around one microsecond. 

Finally, the analysis performed for the pure dust ball collapse has been repeated for the actual scenario of focus -- admitting a (single mode of a) massless scalar field. It has been established in the previous section already, that the global spacetime structure stays the same, which in particular allows us to define and evaluate the entanglement entropy the same way as above. We then expect the results to be perturbations of the case, when the scalar field is absent. The actual calculations have been performed via the same algorithm as in the pure dust case, with the exception that now the function $a(T)$ needed to be determined numerically. The process of finding it reduces, however, to simple numerical integration, thus had not significantly increased the level of difficulty. A generic example of the results of the evaluation of $S_k(u)$ is presented in Figure~\ref{rysSks}. The shape of the entropy curve stays the same, the only difference is a slight increase in the sizes of all the horizons, as the contribution from the scalar field energy adds to the gravitational effect of the dust. There are no qualitative changes to the entropy curve due to the presence of the scalar field. This has unfortunate consequences for the formulated model, which will be discussed in detail in the next section.

\section{Conclusions}

In this work, we attempted to address the effects of quantumness of spacetime geometry on the black hole information loss paradox by evaluating the entanglement entropy for the physical scenario of the Oppenheimer-Snyder dust ball collapse described in the framework inherited from Loop Quantum Cosmology. Our main goal was the determination of the global spacetime structure and evaluating the entanglement entropy in presence of a (single mode of the) massless scalar field in the so-called ``rainbow metric'' scenario, where the backreaction of the metric is taken into account. The exact treatment of the model was a reexamination of the studies in \cite{Parvizi:2021ekr}: the interior of the spherical dust ball has been described in the Loop Quantum Cosmology setting and its dynamics approximated via the effective semiclassical dynamics in the leading (zeroth) order and subsequently glued with two classes of the Vaidya-like metric via standard glueing techniques of classical GR \cite{Poisson:2009pwt}. Imposing the assumption of the existence outside of the dust ball additional translation Killing field, timelike in the regions where $F(X)>0$ (making the exterior metric components depend on the radial coordinate only) allowed to uniquely determine the spacetime geometry via mixed analytical and numerical methods, in particular allowing to write the metric analytically in the absence of the scalar field. 

In both the presence and absence of the scalar field the resulting global spacetime structure qualitatively resembles that of the Reissner–Nordstr\"om space-time (see Fig.~\ref{conformal}), a result that was reported already in the studies of the loop quantum Schwarzschild black hole studied in the midi-superspace framework \cite{Gambini:2008dy, Gambini:2013ooa} (however, authors claim that no singularities are present there). The dust ball features bouncing behaviour characteristic for LQC models, which leads to the (generally expected) black hole to white hole transition picture. The exterior of the dust ball admits two large asymptotically flat regions: past (pre-bounce or black hole) and future (post-bounce or white hole) one. The spacetime features four event horizons: outer and inner ones of the black and white hole respectively. The geometric radius of the outer one is (in the domain of validity of the semiclassical approximations being components of the treatment) just slightly corrected radius of a classical black hole and the inner one is slightly larger than the radius of the dust ball at the bounce. In consequence the studied system realizes the first scenario in the loop black hole paradigm originally discussed in \cite{Ashtekar:2005cj}, where the pre-bounce and post-bounce regions are disconnected -- any causal trajectory between them must pass the region where the quantum geometry effects are strong.

One particular and worrisome feature of the above structure is the presence of the timelike singularity characteristic to Reissner–Nordstr\"om solutions. That singularly persists when the scalar field is included and is reachable from appropriate regions of the future and past null infinity by the null geodesics. This has far-reaching consequences for the subsequent stage of our studies.

For the determined structure of spacetime the entanglement entropy was determined via a slight expansion of the formula found in \cite{Bianchi:2014qua, Bianchi:2014bma} and based on the analysis of the radial null geodesics. The results for both the pure dust and dust with scalar field case are qualitatively similar (see Figs.~\ref{rysS}, \ref{rysSks}). In all cases one can distinguish three regions (epochs) separated by the ``singular'' regions where the null geodesics starting (terminating) at the null infinity reach (originate from) the singularity: 
\begin{enumerate}[(i)]
  \item Black hole, where the null geodesics stay entirely in the causal past of the strong curvature region. There the entropy grows from zero to $+\infty$.
  \item \label{it:quant} The quantum region, where the null geodesics pass between the large asymptotically flat regions through the high curvature region. There the calculations are significantly affected by the quantumness and the discreteness of the geometry. Here, the entanglement entropy can be determined only up to a global additive constant and grows from $-\infty$, reaches a certain finite maximum and again drops to $-\infty$. 
  \item White hole, where the null geodesics stay entirely in the causal future of the strong curvature region. There the entropy (determined again up to an additive constant) grows from $-\infty$ to some finite value.
\end{enumerate}
Remarkably, the epoch of quantum transition (\ref{it:quant}) lasts for a time proportional in the leading order to the collapsing dust ball mass, which for astronomical objects ranges from microseconds to seconds, thus being considerably longer than expected Planck time order. The presence of the scalar field mode does not affect the features above. Its only effect is a slight increase of the horizons sizes and the quantum transition region time length, which is the effect of the scalar field contributing to the black/white hole matter content.

The general picture of the gravitational collapse process and the fate of the black hole information loss paradox following from the studies in this article is rather worrisome. First of all, the removal of the spacelike black hole singularity present in the classical dust collapse scenario has led to the spacetime featuring timelike Reissner-Nordstr\"om-like singularity. In consequence, the entanglement entropy could not be determined for two large portions of the collapsing dust ball evolution. Furthermore, in the limit of approaching the boundaries of these regions (portions of the evolution) the values of the entropy are diverging to infinity. This implies, that the black hole information loss paradox persists in this picture. Not only the collapsing and expanding matter epochs are classically separate (connected only through the region of dominant quantum geometry corrections) but the spacetime is not singularity free.

The persistence of the singularity has far-reaching consequences, potentially putting in question the physicality of the model considered. First of all, in classical GR the Reissner-Nordstr\"om solution is unstable. Upon perturbation, it ``collapses'' into a Schwarzschild-like picture of matter collapse and black hole formation -- the inner horizon of Reissner-Nordstr\"om solution transforms to the singular hypersurface \cite{Simpson:1973ua, Poisson:1989zz}. Therefore, since the field equations are modified by the polymeric quantum nature of the spacetime, the question regarding the stability remains open, however, it provides a strong indication that perturbations may alter the global spacetime structure of the model significantly. Second, the timelike singularity persists, once a small (yet nonlinear) perturbation in the form of a scalar field mode is introduced. In such a case, the neighbourhood of the singularity admits a nontrivial matter content of which energy density reaches infinity. However, the existing experience when extracting the dynamics out of the loop quantized models shows, that generically the loop quantum modification regularizes the singularities (expectation of which has been discussed already in \cite{Thiemann:1997rv}, in particular making the (matter) energy densities and shears finite. While it is not a rigid proof of the incorrectness of the above model, it provides a strong indication, that more advanced treatment would resolve the newly generated singularity. This would however again strongly modify the global structure of studied spacetime. 

Another deficiency of the model is the fact that, while featuring event horizons, it does not incorporate a mechanism of black hole evaporation by Hawking radiation. The matter is reintroduced into the Universe only through the black-to-white hole transition. One can however incorporate the evaporation effects in a phenomenological way applied often in classical GR, assuming a slow change of the dust ball mass due to unspecified interactions with Hawking quanta passing the horizon and considering a solution with slowly varying mass. Applying such reasoning to the black hole region leads to a decrease of the collapsing ball mass. In a naive scenario of trusting the semiclassical approximations used in the construction of this model all the way, the mass would decrease till reaching the extreme case (corresponding to the mass determined via \eqref{eq:Mext}) when the inner horizons coincide with the outer ones. However, one has to take into account, that the derivation of the form of the metric in Sec.~\ref{secexterior} relies in its core on the low-order semiclassical approximation of the loop quantum dust ball Oppenheimer-Snyder collapse. On the genuine quantum level the system is consistent with a model studied in \cite{Husain:2011tm}. In particular, the results regarding the form of the physical Hilbert space and the physical state dynamics can be applied directly here. From there the variances of the state volume, say $\mathcal{V}(T)$, can be easily found for example via the application of methods of \cite{Kowalczyk:2022ajp}. One can in particular find the growth of variance through the bounce analogous to eq.~(5.35) in \cite{Kowalczyk:2022ajp} (see also the analysis of the similar problem in \cite{Kaminski:2010yz}) as well as formulate the analogue of the Heisenberg uncertainty principle involving the (variations of the) volume and the mass of the ball in distant past/future. Given that for near-extremal case the mass is of $m_{{\rm Pl}}$ order, the relative variances of the mass and the volume in at least one epoch will be of the order of the expectation values themselves. In consequence, the state loses all its semiclassicality \cite{unpub}.\footnote{For the same reason of large variances and ill-defined glueing the extremal solution reported in \cite{Lewandowski:2022zce} does not exist, being simply an artefact of pushing the semiclassical approximations well beyond their domain of applicability.} Being a completely spread-out state, it can no longer be glued with a classical metric at the surface of the dust ball, thus rendering the whole procedure inapplicable and its results physically meaningless. As a consequence, once the Hawking evaporation is included on the phenomenological level, only the lower part of the global spacetime structure diagram (Fig.~\ref{conformal}) can be considered reliable. The dust ball forms a black hole that evaporates till at least the horizon neighbourhood is rendered fully quantum and the description presented here no longer applies. 

Given, that the main deficiency of the model is the fact that it still admits singularities (which, as discussed above, indicates its possible unphysicality), let us recall, that the exterior metric has been determined through glueing conditions at the surface of the dust ball, where the interior metric has been determined via LQC methods. In general, the gluing conditions themselves are not sufficient to determine the exterior metric. For that one would need some information about the form of quantum analog of Einstein equations (which is not yet available) and/or additional boundary data. In absence of these, the solution has been fixed uniquely by the condition parachuted from classical dust collapse -- stationarity of the exterior. It is this last condition, which by fixing the solution leads directly to the formation of the timelike singularity. Therefore, we conclude, that in more accurate physical scenario the stationarity will be broken and the spacetime will need to be fully dynamical at least in the region corresponding to the BH$\to$WH transition. This however requires further studies, specifically implementing treatments designed for nonperturbative inhomogeneous scenarios.

\begin{acknowledgments}
  This work was supported in part by the Polish National Center for Science (Narodowe Centrum Nauki -- NCN) grant OPUS 2020/37/B/ST2/03604.
\end{acknowledgments}

\bibliography{biblio}

\end{document}